\documentclass[twocolumn,aps,pre,showpacs]{revtex4}

\usepackage{graphicx}

\usepackage{amsmath}

\usepackage{amssymb}

\usepackage{color}

\begin{document}

\title{Spectral correlations in finite-size Anderson insulators}

\author{T.~Micklitz}

\affiliation{
Centro Brasileiro de Pesquisas F\'isicas, Rua Xavier Sigaud 150, 22290-180, Rio de Janeiro, Brazil 
}

\date{\today}

\pacs{73.20.Fz,73.21.Hb,73.22.Dj}

\begin{abstract}

We investigate spectral correlations in quasi one-dimensional Anderson insulators  
with broken time-reversal symmetry. While energy levels are uncorrelated in the thermodynamic 
limit of infinite wire-length, 
some correlations remain in finite-size Anderson insulators. 
Asymptotic behaviors of level-level correlations in these systems 
are known in the large- and small-frequency limits, 
corresponding to the regime of classical diffusive dynamics and 
the deep quantum regime of strong Anderson localization. 
Employing non-perturbative methods and a mapping to the Coulomb-scattering problem, 
recently introduced by {\it M.~A.~Skvortsov} and {\it P.~M.~Ostrovsky}, we derive a 
closed analytical expression for the spectral statistics in the classical-to-quantum region 
bridging the known asymptotic behaviors. We further discuss how Poisson statistics   
at large energies develop into Wigner-Dyson statistics as the wire-length decreases. 
 
\end{abstract}

\maketitle

\section{Introduction}

The spectral statistics of a quantum mechanical system gives interesting insight into its dynamics. 
It is for example common to uncover integrable or chaotic dynamics  
by establishing Poisson or Wigner-Dyson statistics in the spacing of energy levels~\cite{WDtoP}. 
Universal spectral properties typically emerge if certain scaling limits are applied.  
It seems less appreciated that it is sometimes the non-universal corrections which   
store the more interesting information. 

Low-dimensional, disordered systems of 
non-interacting particles are a prominent example for such a situation.  
At large length-scales the quantum dynamics in such systems 
is dominated by strong Anderson localization.  
Eigenstates then occupy a negligible fraction of the total system's volume, 
and universal Poisson statistics applies in the thermodynamic limit. 
 It predicts 
the absence of correlations
in disorder averages $\langle ...\rangle$ of 
the global density of states $\nu$ at different energies,
 $\langle \nu(\epsilon)\nu(\epsilon+\omega)\rangle
\overset{L\to \infty}{\to }
\langle \nu(\epsilon) \rangle \langle\nu(\epsilon+\omega)\rangle$.
In any finite-size Anderson insulator the situation is  
more interesting. 
Non-universal correlations survive 
and give information
on the system's quantum mechanical dynamics.

Correlations of close-by levels store information on the 
deep quantum regime establishing in the long-time limit.  
The accumulation of quantum interference processes fully 
localizes particles at large time-scales. The remaining  
dynamical processes are
tunneling events between 
 almost degenerate, far-distant eigenstates~\cite{Mott,1dchain,log,Ivanov2012}. 
 Mott's picture of resonant levels gives an intuitive 
 explanation for the spectral 
 correlations in this deep quantum regime: 
The hybridization $\Gamma$
between distant localized states   
decays exponentially on 
the localization length $\xi$.
For a given level-separation $\omega$ 
there is thus a distance, the Mott-scale $l_\omega$, above which $\Gamma$ 
falls below $\omega$.  
Levels separated by energies larger than $\omega$ 
are uncorrelated. Consequently,
in a $d$-dimensional Anderson insulator 
$\langle \nu(\epsilon)\nu(\epsilon+\omega)\rangle
\overset{L\gg\xi}{\to } ( 1- \alpha_d (l_\omega/L)^d )
\langle \nu(\epsilon)\rangle\langle\nu(\epsilon+\omega)\rangle$  
 and the connected correlation function of nearby levels 
is proportional to a power of the Mott-scale,  
$\langle \nu(\epsilon)\nu(\epsilon+\omega)\rangle_{\rm con.} =\alpha_d (l_\omega/L)^d$, 
where $\alpha_d$ some numerical factor.

Correlations at large level-separation $\omega$, on the other hand, give information on 
the dynamics on short time-scales. Quantum interference processes in the short-time 
limit remain largely undeveloped, and 
spectral correlations of far distant levels reflect 
classical diffusion. 

The classical-to-quantum crossover of
spectral correlations in finite-size Anderson insulators 
is unexplored. 
Evidently, this is owed to the fact that the strongly localized regime presents the strong coupling 
limit of the underlying effective field-theory for disordered systems~\cite{EfetovBook,EfetovAdv} 
whose analysis is challenging. 
The situation is similar 
to that previously encountered in fully ergodic {\it chaotic} systems. 
Early on it was conjectured that the classical-to-quantum crossover
of spectral correlations in chaotic systems 
 follows Wigner-Dyson statistics~\cite{BGS}. 
A proof of this `Bohigas-Giannoni-Schmit-conjecture', however, turned out  
 challenging. 
 The reason is a similar as in Anderson insulators: 
 arbitrary orders of quantum interference processes have to be taken into account to 
describe how classical correlations at large energies~\cite{berrydiag} 
evolve into avoided crossings at small energies. 
 In chaotic systems this requires the summation of 
 infinite numbers of periodic orbits and their encounters. 
Progress in this direction
has only been achieved recently~\cite{essen1,essen2,essen3}.

In the present paper we address the classical-to-quantum crossover in the spectral correlations 
of finite-size Anderson insulators. 
Concentrating on
quasi one-dimensional wires 
belonging to the unitary symmetry class 
we derive a closed analytical expression for
the connected level-level correlation function,
\begin{align}
\label{2point}
K(L,\omega)
&=
{\langle \nu(\epsilon)\nu(\epsilon+\omega)\rangle\over
\langle \nu(\epsilon)\rangle\langle \nu(\epsilon+\omega)\rangle}
-
1.
\end{align}
Our result is valid at arbitrary level-separaiotns $\omega$, and thus bridges 
 the known asymptotic behaviors of correlations between close-by and far-distant levels.

The outline of the paper is as follows. Section~\ref{sec:a} reviews the known asymptotic behaviors 
of the level-level correlation function in the classical and deep quantum regimes.
In Sec.~\ref{sec:b} we state our main result, i.e. 
spectral statistics in the classical-to-quantum crossover region,
and explain in detail its derivation.
In Sec.~\ref{sec:c} we present some results for the Poisson-to-Wigner-Dyson crossover 
of level statistics with decreasing wire-length. 
Sec.~\ref{sec:d} summarizes our results. Several technical details are delegated to   
the appendices. Throughout the paper we set $\hbar=1$.

\section{Classical and deep quantum regimes}
\label{sec:a}

In this section we briefly review known asymptotic behaviors of the level-level correlation 
function in Anderson insulators $L\gg\xi$. 

At short time-scales quantum interference processes 
are largely undeveloped. The dynamics is not affected by weak localization corrections and remains 
classically diffusive. 
The leading level-level correlation function, Eq.~\eqref{2point}, in this classical regime 
has been derived by Altshuler and Shklovskii already three decades ago~\cite{AltshulerShklovskii}.
Using diagrammatic perturbation theory they   
found for a quasi one-dimensional geometry 
 \begin{align}
 \label{as}
 K(\omega)
 \propto -{\xi\over L} \left({\Delta_\xi \over \omega}\right)^{3/2},
 \quad
 \omega\gg \Delta_\xi.
 \end{align}
 Here we introduced the localization length $\xi=2\pi \nu D S$,  
  the level-spacing in a localization volume
$\Delta_\xi = (2\pi \nu S\xi)^{-1}$, and $\nu$ is the average DoS. 
$S$ is the wire cross-section and 
$D=v_F^2\tau$ denotes the diffusion constant with $v_F$ the Fermi velocity and $\tau$ the elastic scattering time.
The relevant diagram is shown in the inset of Fig.~\ref{fig1}. Noting that it contains two classical diffusion modes 
$D(q,\omega)\propto1/(q^2+i\omega)$, the $\omega$-dependence of Eq.~\eqref{as} is readily understood 
from simple power-counting $\int_{\sqrt{\omega}} dq/q^4\sim \omega^{-3/2}$.

Eq.~\eqref{as} gives the leading contribution in the small parameter $\Delta_\xi/\omega$. 
Corrections of higher orders ${\cal O}(\Delta_\xi/\omega)$ store information on 
quantum interference processes. These start to become relevant on time-scales 
exceeding the classical regime $t\ll1/\Delta_\xi$. 
In the non-classical regime 
diffusion slows down due to weakly localizing quantum interference processes.  
Accumulation of these processes modifies classical diffusion,    
and localization eventually becomes strong as one approaches  
the Heisenberg-time $t\sim 1/\Delta_\xi$.  

In the strongly Anderson localized regime $t\gg 1/\Delta_\xi$ 
classical diffusion is stopped completely. The remaining dynamical processes 
 in this deep quantum regime are probed by correlations of close-by 
 levels $\omega\ll\Delta_\xi$. Correlation function~\eqref{2point} 
  at small level-separations shows 
 logarithmic level-repulsion~\cite{altlandfuchs,Ivanov2012},
 \begin{align}
 \label{llr}
 K(\omega)
 \propto -{\xi\over L} \log(\Delta_\xi/\omega),
 \quad
 \omega\ll \Delta_\xi.
 \end{align}
Eq.~\eqref{llr} neglects 
corrections smaller in ${\cal O}(\omega/\Delta_\xi)$
and is understood within Mott's picture of resonant levels~\cite{log} 
already mentioned in the introduction. 
Indeed, correlations of nearby levels are due to 
tunneling events between almost degenerate 
states at a distance of the Mott-scale. The physics is  
captured by the two-level Hamiltonian~\cite{altlandfuchs,Ivanov2012} 
\begin{align}
\label{twoRess}
H(\epsilon,\delta\epsilon,x)
&=
\begin{pmatrix}
\epsilon + \delta\epsilon & \Gamma(x) 
\\
\Gamma(x) 
& \epsilon - \delta\epsilon 
\end{pmatrix},
\end{align}
where the hybridization $\Gamma(x)\approx \Delta_\xi e^{-x/\xi} $ 
accounts for a finite overlap of wave-functions centered at distances 
$\xi\lesssim x< L$ (see also Fig.~\ref{fig1}).
Here $\epsilon$ and $\delta\epsilon$
are the mean-level and level-splitting, respectively, which for simplicity are  
both assumed uniformly distributed. 
 Correlation function~\eqref{2point} for the simple model \eqref{twoRess}
reads 
\begin{align}
\label{Mottargument}
K(L,\omega)+1
&=
 \int_\xi^L
  { dx\over L}
\langle 
\delta(\omega-\sqrt{ \delta\epsilon^2+\Delta_\xi^2e^{-{2x\over\xi}} })
\rangle_{\delta\epsilon}, 
\end{align}
where the average 
$\langle...\rangle_{\delta\epsilon}$ is over the level-splitting. 
For close-by levels 
(i.e. $\omega$ smaller than the support of the distribution of $\delta\epsilon$) 
 integral Eq.~\eqref{Mottargument} receives its 
finite contributions from distances $x\gtrsim\ell_\omega\equiv-2\xi\ln(\Delta_\xi/\omega)$, 
larger than the Mott-scale. Subtracting the uncorrelated contribution to the level correlation function 
one finds 
$K(L,\omega\ll\Delta_\xi) \propto \ell_\omega/L$. That is, level-correlations 
are proportional to the configuration space volume for which
hybridization between localized wave-functions is strong enough to result in 
noticeable level-repulsion.

The asymptotic behaviors of the level-level correlation function 
at large and small level-separations, Eqs.~\eqref{as} and \eqref{llr}, 
are well-established for decades~\cite{AltshulerShklovskii,altlandfuchs}. 
The correlation function bridging the classical and deep quantum regimes 
is unknown. 
In the next section we derive 
a closed analytical expression for the latter.

\section{Classical-to-quantum crossover}
\label{sec:b}

We next present our main result, i.e. the level-level correlation function Eq.~\eqref{2point}
in a quasi one-dimensional Anderson insulator. 
We then proceed with a detailed derivation of our result.  
In Sec.~\ref{sec:b1} we introduce the relevant field-theory. 
Sec.~\ref{sec:b2} discusses a mapping from the Anderson 
localization problem to a Coulomb-scattering problem.

\subsection{Results}

We start out from a compact representation of
the level-correlation function Eq.~\eqref{2point}  
in terms of the Green's function for the 
Coulomb-scattering problem,
\begin{align}
\label{KscatPror}
K(L,\omega)
&=  
 -{2\pi \xi \over L}  
 {\rm Re}\,
\lim_{\bold{r}\to\bold{r}_0}
\partial^2_\kappa
G_0(\bold{r},\bold{r}_0),
\end{align} 
applicable for quasi one-dimensional 
Anderson insulators, $L\gg \xi$, within the unitary symmetry class.  
Its derivation is given in the next subsections, where we also introduce  
the Green's function $G_0$ for the non-relativistic $3d$ 
Coulomb-problem~\cite{skvortsov}.
A general closed-form expression of the latter 
has been derived by Hostler half a century ago~\cite{Hostler1,Hostler2},
\begin{align}
\label{gf}
G_0(\bold{r},\bold{r}')
&=
{ (\partial_u-\partial_v)\sqrt{u}K_1(2\sqrt{\kappa u})\sqrt{v}I_1(2\sqrt{\kappa v})
\over 2\pi |\bold{r}-\bold{r}'|}.
\end{align}
Here we have 
introduced $u= r+r'+|\bold{r}-\bold{r}'|$, $v= r+r'-|\bold{r}-\bold{r}'|$
and $K_m$ and $I_m$ are Bessel functions of the first kind.
Inserting the above Green's function~\eqref{gf} into Eq.~\eqref{KscatPror} 
 gives the level-level correlation function in a finite-size Anderson insulator,  
\begin{align}
\label{Kfin}
K(L,\omega)=
-{16\xi \over L }
 {\rm Re} \,
 {\cal K}(4\omega/i\Delta_\xi),
\end{align}
where
\begin{align}
\label{kfin}
 {\cal K}(z)
&=2\partial_z K_0(\sqrt{z}) I_0(\sqrt{z})
\nonumber \\
&=  
(1/\sqrt{z})
\left(
K_0(\sqrt{z})I_1(\sqrt{z})
-
K_1(\sqrt{z})I_0(\sqrt{z})
\right).
\end{align}

\begin{figure}[t]
\begin{center}
\includegraphics[width=8.6cm]{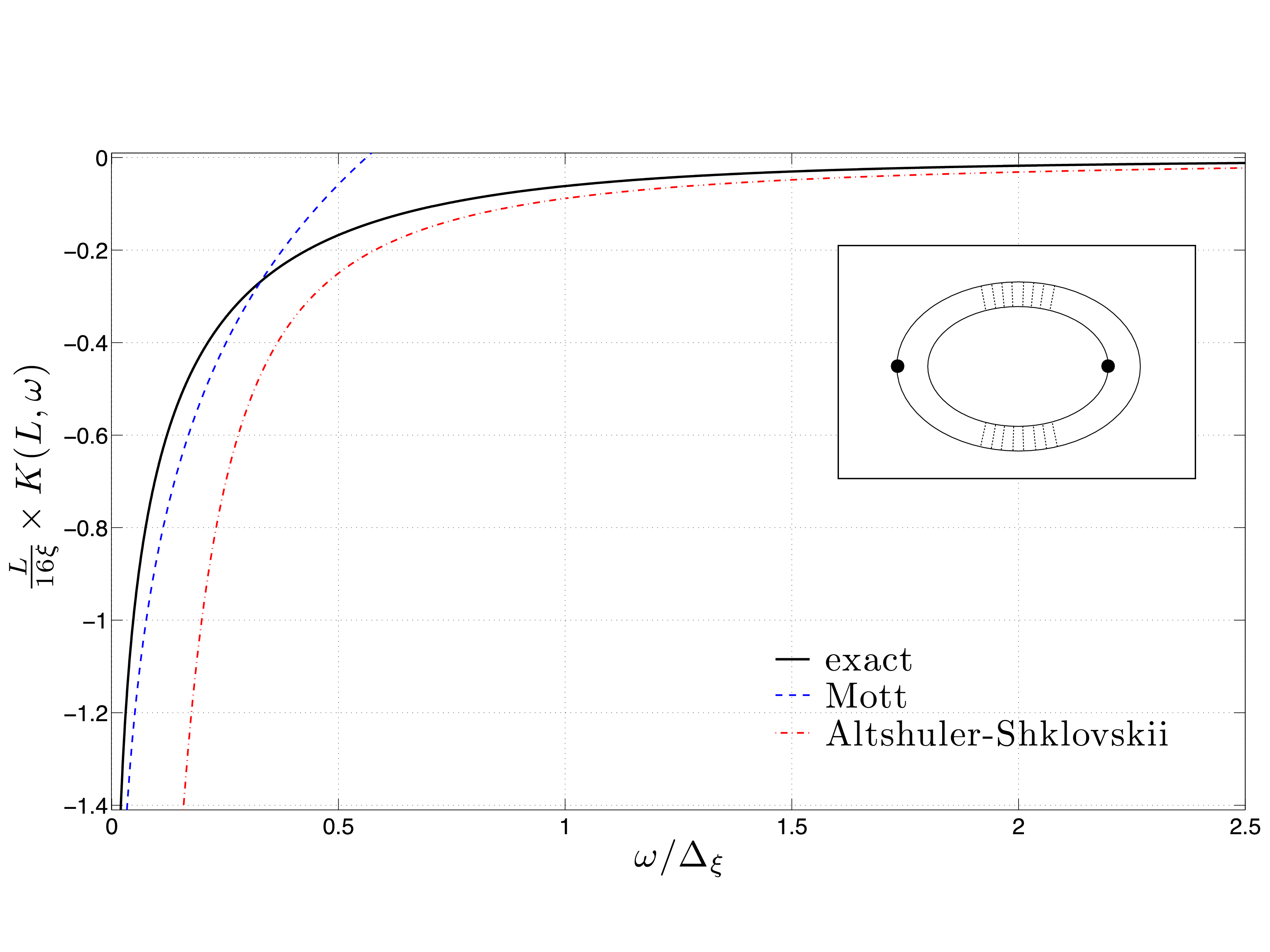}
\end{center}
\vspace{-15pt}
\caption{
Spectral correlations in a finite-size Anderson insulator, Eqs.~\eqref{Kfin} 
and \eqref{kfin}. For reference we also show  
previously known limits (dashed lines), 
i.e.
 $K(L,\omega)
= 
 -(4 \xi/ L) \left( \Delta_\xi/2\omega\right)^{3/2}$ 
at large,
and 
 $K(L,\omega)
= 
- (8 \xi /L )
\left[
\log\left( \Delta_\xi / \omega \right)
-
2\gamma
 \right]$ 
at small frequencies respectively, see Eq.~\eqref{asympt}.  
Inset: Feynman diagram accounting for Altshuler-Shklovskii correlations 
Eq.~\eqref{as}; dashed lines represent classical diffusion modes.
}
\label{fig1}
\end{figure}

Eqs.~\eqref{Kfin} and \eqref{kfin} are the main result of this paper.
It describes how the Altshuler-Shklovskii correlations in the classical regime 
evolve into logarithmic level repulsion in the deep quantum regime. 
Poisson statistics only applies in the thermodynamic limit
where the residual correlations in Anderson insulators 
vanish $K(L\to \infty,\omega)=0$. 
From Eq.~\eqref{kfin} we readily extract the asymptotic correlations 
of far-distant and close-by levels~\cite{Gradsteyn},
\begin{align}
\label{asympt}
K(L,\omega)
&= 
-{ 4\xi \over L } 
\begin{cases}
\left( {\Delta_\xi\over 2\omega}\right)^{3\over2}
+
{3\over 32} \left({\Delta_\xi\over 2\omega}\right)^{5\over2}
+ ...,
\, \, \,
 \omega \gg \Delta_\xi,
\\
2\log\left( { \Delta_\xi \over \omega} \right)
-
4\gamma
+
{3\pi \omega\over 2\Delta_\xi}
+ ...,
\,
\omega \ll \Delta_\xi.
\end{cases}
\end{align}
Here $\gamma\simeq0.577$ is the Euler-Mascheroni constant.
The spectral correlation function, Eqs.~\eqref{Kfin} and \eqref{kfin}, is shown in Fig.~\ref{fig1}. 
For reference we also display the previously known asymptotic behaviors.

\subsection{ Field-theory}
\label{sec:b1}

Our derivation of representation \eqref{KscatPror} 
starts out from a field-theory description of 
the level-level correlation function~\cite{EfetovBook,altlandfuchs}, 
\begin{align}
\label{KsM}
K(L,\omega)
&=
-{{\rm Re}\over 32} 
\int_0^L dx \int_0^L dx' \,
\langle 
P_+[Q(x)] 
P_-[Q(x')] 
\rangle_Q, 
\nonumber
\\
P_\pm[Q] 
&=
{\rm str}\left( (Q-\Lambda)(1\pm\Lambda)k\right). 
\end{align}
Here  the average is with respect to the diffusive nonlinear 
$\sigma$-model action introduced below,
\begin{align}
\langle...\rangle_Q 
&=
\int {\cal D}Q\, e^{S_\sigma[Q]}.
\end{align} 
Integration $\int {\cal D}Q$ is over $4\times 4$ supermatrices 
$Q(x)=\{Q^{\alpha\alpha'}_{ss'}(x)\}$ 
 from the supergroup $U(1,1|2)$
 obeying the nonlinear constraint $Q^2(x)=\openone$. 
Diagonal $2\times 2$ upper- left and lower-right matrix blocks $Q^{\mathrm{bb}}$ and $Q^\mathrm{ff}$, respectively, 
contain complex numbers. 
The off-diagonal blocks $Q^\mathrm{bf,fb}$ consists of Grassmann variables. 
The subscript indices $Q_{ss'}$, $s,s'=\pm$ discriminate between `retarded' and
 `advanced' components of the matrix field. 
 The $c$-number content of the $Q$-field manifold 
 reduces to the direct product of the hyperboloid $U(1,1)/[U(1)\times U(1)]$
  in the ${\rm bb}$-block and the sphere $U(2)/[U(1)\times U(1)]$ in the ${\rm ff}$-block.   
It is parametrized by non-compact 
and compact 
variables,  
$1\leq \lambda_{\rm bb}$ 
and 
$-1\leq \lambda_{\rm ff}\leq 1$, 
respectively.
The matrix $\Lambda=\{s\, \delta_{ss'}\}$ is the identity matrix in boson-fermion space 
but breaks symmetry in advanced-retarded space. 
 $k=\Lambda \sigma_3^{\rm bf}$ is a diagonal matrix 
with 
$\sigma_3^{\rm bf}$ 
breaking also symmetry in boson-fermion space, and 
`${\rm str}$' is the generalization of the matrix trace to graded space. 

The diffusive nonlinear $\sigma$-model action reads~\cite{EfetovBook,EfetovAdv},
\begin{align}
\label{action}
S_\sigma[Q]
&=  -{\pi\nu S\over 4} \int dx\, 
{\rm str}\left( 2i\omega Q \Lambda
+
D(\partial_x Q)^2 \right).
\end{align}
It is the low-energy effective field-theory for 
Anderson localization, here 
for a quasi one-dimensional 
geometry and in the presence of a weak time-reversal symmetry breaking vector potential. 
We refer for a derivation of action \eqref{action} to Refs.~\cite{EfetovBook,EfetovAdv}, 
and here point out its 
structural similarity to the Hamilton function of a classical ferromagnet 
in an external magnetic field. 
The latter breaks rotational invariance of the exchange interaction, and
a similar role is played by 
the potential $V(Q)\sim {\rm str}\left( Q \Lambda \right)$ in the $\sigma$-model action.
$V(Q)$ breaks 
the invariance of the kinetic term 
 $K(Q)\sim{\rm str}\left( [\partial_xQ]^2\right)$ 
 under general rotations $U(1,1|2)$.
 The strength of symmetry breaking 
is given by the level-separation $\omega$.  
Energies $\omega\gg\Delta_\xi $ larger than the level-spacing
imply strong symmetry-breaking. 
$Q$-fields are then 
pinned to the mean-field $\Lambda$ and small fluctuations
can be accounted for perturbatively. 
A straightforward perturbative expansion at large energies 
gives the leading correlation 
function  $K(\omega\gg \Delta_\xi)\propto \omega^{-3/2}$ discussed 
in the previous section. 
Once $\omega \lesssim \Delta_\xi$ falls below the level-spacing, fluctuations become large   
 acting to restore the full symmetry of the kinetic term. 
A direct integration of Eq.~\eqref{KsM} is hindered by the presence of 
the rotational symmetry breaking potential $V(Q)$.   
Non-perturbative methods, discussed below, 
have to be applied to address the low energy correlations 
$\omega \lesssim \Delta_\xi$.

\subsection{Non-perturbative solution}
\label{sec:b2}

We next derive the alternative representation 
of the correlation function~\eqref{KsM} 
 in terms of 
the Green's function for the Coulomb scattering problem~\cite{skvortsov}. 
To this end we recall that
one can map the integral Eq.~\eqref{KsM} to a set of equivalent differential 
equations~\cite{efetovreview,EfetovBook}.  
Indeed, identifying $Q$ with the coordinate of a multi-dimensional quantum
particle and the wire-coordinate with time, Eq.~\eqref{action}  
reads as the Feynman path-integral of a particle with kinetic energy $K(Q)$ 
moving in the potential 
$V(Q)$. Alternatively to calculating the path-integral, 
one can solve the corresponding `Schr\"odinger equation', 
known as transfer-matrix equations.

Similar to a radial potential in quantum mechanics, the high degree of symmetry of the potential
($V$ is invariant under similarity transformations of $Q$ leaving $\Lambda$ invariant) 
reduces the effective dimensionality of the problem. As detailed in Appendix~\ref{app1},
  correlation function \eqref{KsM} reduces to an integral over the $c$-number  
  variables $\lambda_{\rm bb}$ and $\lambda_{\rm ff}$. 
The integrand is expressed in terms of a ground- and an excited-state 
  wave-function of the underlying Schr\"odinger 
  equation. As the Laplace-operator on the 
  $Q$-matrix manifold has a rather complex structure  
 closed solutions of the latter are not available. 
    
    Progress in this direction has been made in a 
    recent work by Skvortsov and Ostrovsky~\cite{skvortsov}. 
  There the authors elaborate on a connection between 
  localization in quasi one-dimensional systems within the unitary class
  to scattering in a Coulomb potential. Changing from angle to `Coulomb'-coordinates, 
\begin{align}
 \lambda_{\rm ff} &=(r-r_1)/2,
 \quad
 \lambda_{\rm bb}=(r+r_1)/2,
 \end{align}
 a major simplification occurs when the latter are understood
 as elliptic coordinates of a three-dimensional problem with cylindrical symmetry, 
\begin{align}
 r&=\sqrt{z^2+\rho^2},
 \quad
 r_1=\sqrt{(z-2)^2+\rho^2}.
 \end{align}
 Here $(\rho,\varphi,z)$ are the usual cylindrical coordinates. 
 Following the outlined procedure (and leaving details to Appendix~\ref{app2}) 
the correlation function \eqref{KsM} becomes 
 \begin{align}
\label{kr}
K(L,\omega)
&= {\xi^2 {\rm Re}\over 2\pi L^2}
 \int_0^{ L\over \xi} dt'
\int {d\bold{r}\over r} 
\Phi_0(\bold{r},L/\xi-t') 
\Phi_1(\bold{r} ,t'),
\end{align} 
where $\Phi_0$ and $\Phi_1$ are the ground- and excited-state wave-functions, respectively. 
The latter are solutions of the following transfermatrix equations in Coulomb-coordinates,
\begin{align}
\label{seqcc1}
\left(
 {1 \over  r_1r } \partial_t - 
\partial^2_\bold{r} +  {2\kappa\over r}
\right) \Phi_0(\bold{r},t)
&=0,
\\
\label{seqcc2}
\left(
  {1 \over  r_1r } \partial_t - 
\partial^2_\bold{r} +  {2\kappa\over r}
\right)  \Phi_1(\bold{r},t)
&= {1\over r} \Phi_0(\bold{r},t).
\end{align} 
Here $\partial_{\bold r}$ is the three-dimensional Laplace operator, 
$\kappa=\omega/(4i\Delta_\xi)$ and $t=x/\xi$.  
Eqs.~\eqref{seqcc1} and \eqref{seqcc2} are supplemented by the 
boundary conditions  
$\Phi_0(\bold{r},0)=1/r_1$ and $\Phi_1(\bold{r},0)=0$.
In the strongly localized regime of interest 
homogeneous solutions of the transfermatrix equations
give the leading contribution to Eq.~\eqref{kr}. 
Indeed, inhomogeneous solutions decay exponentially from the boundaries 
and their contributions to the integral Eq.~\eqref{kr} can be neglected for $L\gg\xi$.
Dropping $t$-dependencies of the wave-functions,
Eqs.~\eqref{seqcc1} and \eqref{seqcc2} become spherical symmetric. The 
corresponding boundary conditions read  
$\Phi_0(\bold{r}_0)=1/r_1$ and $\Phi_1(\bold{r}_0)=0$ with $\bold{r}_0=(0,0,2)^t$. 
The reduction to a problem of higher degree of symmetry is a key simplification 
 which allows for an analytical calculation of \eqref{kr} to leading order 
in the small parameter $\xi/L$.

Following Ref.~\onlinecite{skvortsov} we introduce 
the zero-energy Green's function for the non-relativistic $3d$ 
Coulomb-problem,
\begin{align}
\left( 
 \partial^2_\bold{r} -  {2\kappa\over r}
\right) G_0(\bold{r},\bold{r}')
&=\delta(\bold{r}-\bold{r}').
\end{align}
Imposing usual boundary condition 
$G_0(\bold{r},\bold{r}')\overset{\bold{r}\to\bold{r}'}{\mapsto}
\delta(\bold{r}-\bold{r}')$  
the homogeneous 
solution to Eq.~\eqref{seqcc1} with required boundary condition 
affords the representation~\cite{skvortsov}
\begin{align}
\label{phi0}
\Phi_0(\bold{r})= -4\pi G_0(\bold{r},\bold{r}_0).
\end{align} 
Similarly, it is verified that the convolution
\begin{align}
\label{phi1}
\Phi_1(\bold{r})= -\int {d\bold{r}'\over r'} G_0(\bold{r},\bold{r}') \Phi_0(\bold{r}')
\end{align}
is a $t$-independent solution of Eq.~\eqref{seqcc2} with the required boundary condition.
Inserting above solutions \eqref{phi0} and \eqref{phi1} into \eqref{kr} 
one confirms that~\cite{fn1} 
\begin{widetext}
\begin{align}
\label{KscatPro}
K(L,\omega)
&=  
-{8\pi \xi\over  L} 
  {\rm Re}\,
\int d\bold{r} \int d\bold{r}' \,
G_0(\bold{r}_0,\bold{r}){1\over r}G_0(\bold{r},\bold{r}'){1\over r'}G_0(\bold{r}',\bold{r}_0)
=  
 -{2\pi \xi \over L}  
 {\rm Re}\,
\lim_{\bold{r}\to\bold{r}_0}
\partial^2_\kappa
G_0(\bold{r},\bold{r}_0).
\end{align} 
\end{widetext}
This completes our derivation of the
spectral correlation function \eqref{2point} 
in terms of the 
Green's function for the Coulomb-scattering problem.

\section{From Poisson to Wigner-Dyson statistics}
\label{sec:c}

We next address how spectral correlations in a finite-size Anderson insulator
 turn into Wigner-Dyson correlations as the wire-length is reduced.
We recall that in the fully ergodic quantum dot-limit  
level correlations follow
Wigner-Dyson statistics 
$K(L\to 0,\omega)=-(\Delta/\omega \pi)^2\sin^2\left(\omega\pi/\Delta\right)$. 
Here $\Delta = (\nu SL)^{-1}$ is the level spacing of the wire of length $L$. 
Noting that $\Delta/\Delta_\xi \sim \xi/L$,
 one can thus study how Altshuler-Shklovskii correlations at 
 $\omega\gg\Delta_\xi$ evolve 
  into Wigner-Dyson correlations as the wire-length decreases.

\begin{figure}[b]
\begin{center}
\includegraphics[width=8.2cm]{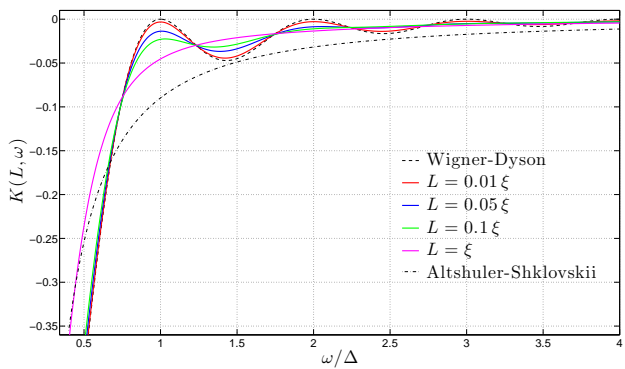}
\end{center}
\vspace{-15pt}
\caption{
Level-level correlations Eqs.~\eqref{k2lo} and~\eqref{hkappa} 
for different ratios $L/\xi$ in the Poisson to Wigner-Dyson crossover region. 
For comparison we also show Wigner-Dyson and Althsuler-Shklovskii correlations 
(dashed and dashed-dotted lines).
}
\label{fig2}
\end{figure}

Correlations  
at arbitrary ratios $L/\xi$ can be 
derived from the inhomogeneous transfermatrix 
equations~\eqref{seqcc1} and \eqref{seqcc2}. 
For levels separated by  $\omega\gg\Delta_\xi$ 
the potential $V$ pins the wave-functions 
to the region $r_1\ll 1$ enforced by the boundary conditions. 
We can thus approximate for the ground-state wave-function 
\begin{align}
\label{largew}
\left(
 {1 \over 2r_1 }
 \partial_t - 
 \partial^2_\bold{r} +  \kappa
\right) \Phi_0(\bold{r},t)
&=0.
\end{align}
The radial symmetry of Eq.~\eqref{largew} substantially simplifies the problem.  
Starting out from the ansatz 
$\Phi(\bold{r},t)= e^{-F(\kappa,t)r_1}/r_1$ 
the function $F$ satisfies   
$ \partial_t F + 2 F^2 - 2 \kappa=0$.
Employing the boundary condition $F(\kappa,0)=0$ one then finds 
\begin{align}
\label{sollo}
 \Phi_0(\bold{r},t)
 &=
 {1\over r_1} e^{-\sqrt{\kappa}\tanh\left(2\sqrt{\kappa}t\right)  r_1}.
\end{align}
Eq.~\eqref{sollo} interpolates between 
the known limits at small and large ratios $L/\xi$. 
Indeed,  
 $\tanh(2\sqrt{\kappa}t)$ approximates to $1$ and $2\sqrt{\kappa}t$
  for $\omega/\Delta$ much larger, respectively, smaller than $\xi/L$. 
This reflects the typical decaying and oscillating behaviors of the ground-state 
 wave-function in the two limits.
 A similar calculation, detailed in Appendix~\ref{app3},
  gives for the level-level correlation function 
  \begin{align}
 \label{k2lo}
 K(L,\omega)
&= 
{ \Delta \xi\, {\rm Im}\over \pi L }\,
 \int_0^{L\over\xi} dt'  \,
\left(
1- e^{-H^{-1}(\omega/i\Delta_\xi,L/\xi,t')} 
\right) 
\nonumber\\
& \qquad \qquad \qquad \times
 \partial_\omega H(\omega/i\Delta_\xi,L/\xi,t'),
 \end{align}  
with 
 \begin{align}
\label{hkappa}
H(z,t,t')
&=
{ \cosh(\sqrt{z}t') \cosh(\sqrt{z}(t'-t))
\over
 \sqrt{z} \sinh(\sqrt{z}t)}.
 \end{align}
We emphasize that Eqs.~\eqref{k2lo},~\eqref{hkappa}
hold for $\omega\gg \Delta_\xi$ and arbitrary ratios $L/\xi$.

The analytical result~\eqref{k2lo},~\eqref{hkappa} 
is displayed in Fig.~\ref{fig2}. It  
 shows how the Altshuler-Shklovskii correlations in long wires 
evolve into the Wigner-Dyson correlations as one approaches 
the quantum-dot limit. 
Unfortunately, we do not know the corresponding result for   
correlations between levels separated by $\omega\lesssim\Delta_\xi$. 
The transfermatrix equations in the 
Poisson-to-Wigner-Dyson crossover region then lack rotational 
symmetry and we were not able to find analytical solutions.

\section{Summary and discussion}
\label{sec:d}

In this paper
we have derived the leading level-level correlations in quasi one-dimensional 
Anderson insulators with broken time-reversal symmetry.
While energy levels are uncorrelated in the thermodynamic limit 
correlations remain in finite-size Anderson insulators. 

The correlations of far-distant and nearby levels reflect the dynamics 
in the classical diffusive regime and the deep quantum regime of 
strong Anderson localization.
They have been well-established for decades~\cite{AltshulerShklovskii,altlandfuchs}. 
This paper discusses the previously unknown 
correlations at  arbitrary level-separations.  
Specifically, our result describes how 
Altshuler-Shklovskii correlations  
at large separations 
turn into logarithmic level-repulsion at small separations. 
Only in the limit of infinite wire-length the correlations vanish, in accordance with 
universal Poisson statistics expected in the non-ergodic system. 

We further discuss how Altshuler-Shklovskii correlations in Anderson insulators turn into 
Wigner-Dyson correlations with  decreasing wire-length. 
A corresponding analysis for correlations between close-by levels remains an open problem.
 
 Finally, we would like to put our results into the context of previous works. 
Spectral correlations of quasi one-dimensional disordered systems
in the Wigner-Dyson-to-Poisson crossover 
have been addressed in Ref.~\onlinecite{altlandfuchs}.
From numerical solutions of the relevant 
transfermatrix equations a qualitative understanding of the 
level-level correlation function 
in the Wigner-Dyson-to-Poisson crossover was obtained. 
Local correlations in the 
density of states within a localization volume were derived 
in Ref.~\onlinecite{skvortsov}. 
The authors find the exact 
ground-state wave-function of 
the homogeneous transfermatrix equation by
mapping the equation to the 
Coulomb-scattering problem. 
It is shown that correlations of different eigenfunctions 
are different in quasi- and strictly one-dimensional geometries~\cite{1dchain}. 
Correlation functions for the {\it global} density of states (discussed in this work) 
and the {\it local} density of states (discussed in Ref.~\onlinecite{skvortsov}) 
both have representations in terms of the 
Coulomb Green's function. 
A similar relation has been observed in 
Ref.~\onlinecite{altshulerlg} in the context 
of parametric correlations, where it  
was connected to a symmetry in the $\sigma$-model. 
Ref.~\onlinecite{Ivanov2009} derives a perturbative expansion for the local density of states 
correlation function
at small frequencies. Their general expression e.g. reproduces statistics of single 
localized wave-functions and predicts re-entrant behavior at the Mott scale 
(similar to that in strictly $1d$ chains~\cite{1dchain}). 
It would be interesting to investigate 
how the findings reported in the present work are obtained within this approach. 
Furthermore, extensions of the discussed results to other symmetry classes remains an 
open problem.

\acknowledgements 
I thank A.~Altland for useful discussions during his visits within 
the program ``Science Without Borders" of CNPq (Brazil)
and for helpful comments on the manuscript.  
Financial support by Brazilian agencies CNPq and FAPERJ are acknowledged.

\begin{appendix}

\section{Transfermatrix equations}
\label{app1}

For self-consistency we state 
the level-level correlation function in a quasi one-dimensional 
wire. In the unitary symmetry class the latter
depends on the $c$-number variables $\lambda_{\rm bb}$, $\lambda_{\rm ff}$.
Following Refs.~\onlinecite{EfetovBook,efetovreview,altlandfuchs} 
one finds 
\begin{align}
K(L,\omega)
&= {\xi^2 \over L^2}
{\rm Re}
\int_0^{L\over\xi} dt'
\int_1^\infty {d\lambda_1} 
\int_{-1}^1 {d\lambda\over \lambda_1-\lambda} 
\nonumber \\
& \qquad  
\times
\,Ê
\Psi_0(\lambda_1,\lambda,L/\xi-t') 
\Psi_1(\lambda_1,\lambda,t').
\end{align}
For notational convenience we  here introduced  
$\lambda_{\rm bb}\equiv \lambda_1$ and $\lambda_{\rm ff}\equiv \lambda$, 
and $\Psi_0$ and $\Psi_1$ are the ground-state and an 
excited-state wave-function, respectively.
The latter follow the transfermatrix equations 
\begin{align}
\left( \partial_t + 2\hat{H}_0 \right) \Psi_0(\lambda_1,\lambda, t) 
& = 0,
\\
\left(\partial_t  +  2\hat{H}_0  \right) \Psi_1 (\lambda_1,\lambda,t) 
&= (\lambda_1-\lambda) \Psi_0(\lambda_1,\lambda,t),
\end{align} 
where we introduced $t=x/\xi$.
The Hamilton operator reads 
$\hat{H}_0=\Delta_Q+\hat{V}$, where
\begin{align}
\Delta_Q
& =
-
{(\lambda_1-\lambda)^2\over 2}
\left(
 \partial_\lambda 
 {1-\lambda^2 \over (\lambda_1-\lambda)^2} \partial_\lambda
+
 \partial_{\lambda_1} 
 {\lambda_1^2-1 \over (\lambda_1-\lambda)^2} \partial_{\lambda_1}
 \right)
\end{align}
is the Laplace operator on the $Q$-field manifold, and
\begin{align}
\hat{V}
&=
- {i\omega \over 4\Delta_\xi}(\lambda_1-\lambda),
\end{align}
is the symmetry-breaking potential. 
The above equations 
should 
be solved 
 with boundary conditions 
for an open wire,
\begin{align}
\Psi_0(\lambda_1,\lambda,0) =1, \qquad
\Psi_1(\lambda_1,\lambda,0) =0. 
\end{align}

\section{Cylindrical coordinates}
\label{app2}

Employing elliptic coordinates 
($\lambda_{\rm bb}\equiv \lambda_1$, $\lambda_{\rm ff}\equiv \lambda$) 
\begin{align}
 \lambda &=(r-r_1)/2,
 \quad
 \lambda_1=(r+r_1)/2,
 \\
 r&=\sqrt{z^2+\rho^2},
 \quad
 r_1=\sqrt{(z-2)^2+\rho^2},
 \end{align}
the Hamilton operator takes the form
\begin{align}
\hat{H}_0
&=-{r_1^2r\over 2}
\left[
\partial_z^2 + \partial_\rho^2 +{1\over \rho}\partial_\rho 
+
{i\omega\over 2\Delta_\xi r}
\right] {1\over r_1}.
\end{align}
Notice that the differential operator 
 is the usual Laplacian in cylindrical coordinates $(\rho,\varphi,z)$, 
acting on cylindrical symmetric functions. 

It is then convenient to make the ansatz $\Psi_i=r_1\Phi_i$
and to express the corresponding Schr\"odinger equations 
in three-dimensional coordinates
\begin{align}
\left(
 \partial_t - 
  r_1r 
  \partial^2_\bold{r} +  2r_1\kappa
\right) \Phi_0(\bold{r},t)
&=0,
\\
\left(
 \partial_t -   r_1r \partial^2_\bold{r} + 2r_1\kappa
\right)  \Phi_1(\bold{r},t)
&= r_1 \Phi_0(\bold{r},t).
\end{align} 
Here $\kappa=\omega/(4i\Delta_\xi)$ and we  
recall that the boundary conditions read 
$\Phi_0(\bold{r},0)=1/r_1$
and $\Phi_1(\bold{r},0)=0$. 
The integration measure transforms into the  
three-dimensional volume element of cylindrical symmetric functions, 
\begin{align}
d\lambda \wedge d\lambda_1 
&=
{\rho\over rr_1} 
dz \wedge d\rho   
\mapsto
{ d\bold{r} \over 2\pi rr_1}, 
\end{align}
where $d\bold{r}=dz \wedge \rho  d\rho \wedge d\varphi$.
The level-level correlation function thus takes the form 
\begin{align}
\label{appcf}
K(L,\omega)
&= { \xi^2  {\rm Re} \over  2\pi L^2}
\int_0^{L\over\xi} dt'
\int {d\bold{r}\over r} 
\Phi_0(\bold{r},L/\xi-t') 
\Phi_1(\bold{r} ,t').
\end{align}

\section{Poisson-to-Wigner-Dyson crossover}
\label{app3}

Level-level correlations  
in systems of arbitrary ratios $L/\xi$ 
can be derived from the inhomogeneous transfermatrix equations.
For levels separated by $\omega\gg\Delta_\xi$
the potential $V$ pins the ground state wave-function 
to the region $r_1\ll 1$ enforced by the boundary condition. 
We may thus approximate Eq.~\eqref{seqcc1} by
\begin{align}
\label{aplargew} 
\left(
 \partial_t - 
 2r_1 \partial^2_\bold{r} +  2r_1 \kappa
\right) \Phi_0(\bold{r},t)
&=0.
\end{align}
Eq.~\eqref{aplargew}
with boundary condition
$\Phi_0(\bold{r},0)=1/r_1$
is solved by
$\Phi_0(\bold{r},t)
 = e^{-\sqrt{\kappa}\tanh\left(2\sqrt{\kappa}t\right)  r_1}/ r_1$. 
Similarly, one may verify that for $\omega\gg\Delta_\xi$ 
the excited-state wave-function
\begin{align}
\Phi_1(\bold{r},t)
&= 
-{1\over 2}\partial_\kappa \Phi_0(\bold{r},t)
\end{align}
satisfies transfermatrix equation 
\begin{align}
\left(
 \partial_t 
 - 
2r_1\partial^2_\bold{r} + 2r_1 \kappa
\right)  \Phi_1^0(\bold{r},t)
&= r_1 \Phi_0(\bold{r},t),
\end{align} 
with  boundary condition $\Phi_1(\bold{r},0)=0$. 
Inserting $\Phi_0$ and $\Phi_1$ into the level-level correlation function~\eqref{appcf} 
one arrives at
Eqs.~\eqref{hkappa} and \eqref{k2lo} stated in the main text.
Notice that in the quantum-dot limit 
$\Phi_0(\bold{r},t)
 = e^{-2\kappa r_1 t }/ r_1$ and 
 $\Phi_1(\bold{r},t)
 = te^{-2\kappa r_1 t }$.
This results in the Wigner-Dyson correlations 
 $K(L,\omega)=-(\Delta/\omega\pi)^2\sin^2(\omega\pi/\Delta)$ applicable  
 at arbitrary ratios $\omega/\Delta$. 
 That is, the restriction   
 $\omega\gg\Delta_\xi\sim \Delta L/\xi$ becomes irrelevant
in the limit $L\ll\xi$.

\end{appendix}


\begin{thebibliography}{99}
\bibitem{fn1}
We anticipate symmetry in the arguments of the Green's function, 
$G_0(\bold{r},\bold{r}')=G_0(\bold{r}',\bold{r})$, evident in Eq.~\eqref{gf}.
\end{thebibliography}
\end{document}